# Design Guidelines for In-line X-ray Inspection in Advanced Packaging Technology: A CoWoS Case Study

Katayoon Yahyaei, M Shafkat M Khan, Guillermo Cadima, Himanandhan Reddy Kottur, and Navid Asadizanjani
Dept. of Electrical and Computer Eng., University of Florida
5000 Malachowsky Hall, 1889 Museum Road
Gainesville, FL 32611 USA
Email: ka.yahyaei@ufl.edu, nasadi@ece.ufl.edu

## Abstract

The shift towards advanced packaging technologies, including 2.5D and 3D integration, addresses the limitations of traditional methods while meeting increasing demands for performance, miniaturization, and efficiency. These methods enhance functionality and support heterogeneous integration but also introduce metrology challenges due to complex, three-dimensional structures. X-ray imaging, crucial for non-destructive inspection, faces compatibility issues such as material density similarities and noise scattering. To address these challenges, we propose a framework based on AI-integrated Design of Experiment (DoE) to develop design guidelines to optimize X-ray compatibility during the design stage. This framework, demonstrated through a case study on Chip-on-Wafer-on-Substrate (CoWoS) packaging, systematically analyzes design parameters and material properties to develop guidelines for improved inspection accuracy. Our method integrates AI to predict outcomes and optimize processes, ensuring high-quality X-ray images and enhancing defect detection. Implementing these guidelines can significantly improve inspection accuracy and reliability, reducing production costs and supporting the efficiency and scalability of advanced semiconductor technologies.

## I. Introduction

The adoption of advanced packaging technologies, including 2.5D and 3D integration, addresses the limitations of legacy packaging techniques while meeting the increasing demands for improved performance, further miniaturization, and enhanced efficiency [1]. The complexity of 3D structures, stacked dies, and varying materials in advanced packaging introduces significant metrology challenges [2]. Defect detection, especially for critical components like through-silicon vias (TSVs) and microbumps, requires precise and robust inspection strategies to identify flaws that may impact connectivity and performance [3]. Moreover, application-specific requirements for in-line inspection significantly impact inspection efficiency, affecting time and computational budgets. While consumer-level devices may accept lower reliability targets, critical applications like automotive systems require 100% sampling of around 1 million units per year for physical inspection [4]. These requirements place limitations on the capabilities of X-ray systems within the in-line inspection workflow, impacting critical parameters such as field-of-view and resolution. These parameters play a pivotal role in determining the quality of X-ray scans. For instance, as illustrated in Figure 1, the resolution difference between 1.28 µm and 0.7 µm in X-ray computed tomography (CT) of an Nvidia Tesla P100 sample, which is designed with TSMC 2.5D Chip-on-Wafer-on-Substrate (CoWoS) technology, significantly influences the scan's detail and effectiveness.

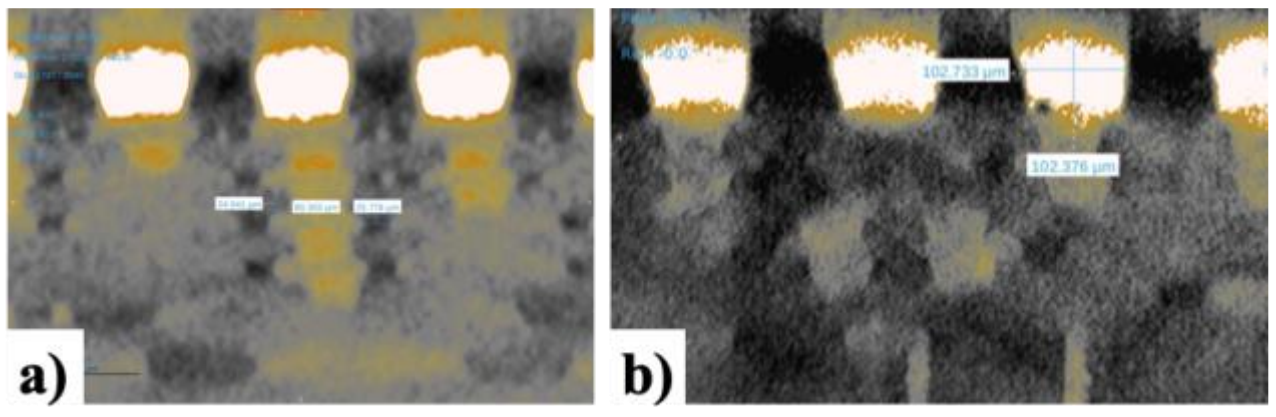

*Figure 1: C4 bumps and substrate cross section of Nvidia Tesla P100 package with a) 1.28 µm and b) 0.7 µm resolution.*

X-ray inspection plays an essential role in advanced package inspection, providing non-destructive, high-resolution capabilities to evaluate high aspect ratio features, detect hidden or buried defects, and examine complex material layers [5]. However, X-ray inspection challenges, also called X-ray compatibility issues, occur due to factors like similar material densities, noise scattering, and the presence of layered structures, which can obscure defects, distort measurements, and increase inspection time and costs [6]. An example of advanced packaging features that pose challenges for resolution in X-ray imaging, particularly in in-line inspections, is the redistribution layers (RDLs) in interposers. Figure 2-a depicts a 0.7 µm X-ray CT scan of an overall cross-section of an Nvidia Tesla P100 sample, highlighting an area that includes the die, interposer, and substrate. Figure 2-c provides a detailed view of the region encircled in red in Figure 2-b. The red rectangle in Figure 2-c indicates the location of the RDLs beneath the microbumps within the interposer. However, due to the intricate structure of this area, the RDLs cannot be adequately resolved in the X-ray image. Another example of X-ray imaging challenges is the beam hardening imaging artifact that usually occurs around C4 bumps due to their large size and high-density material composition.

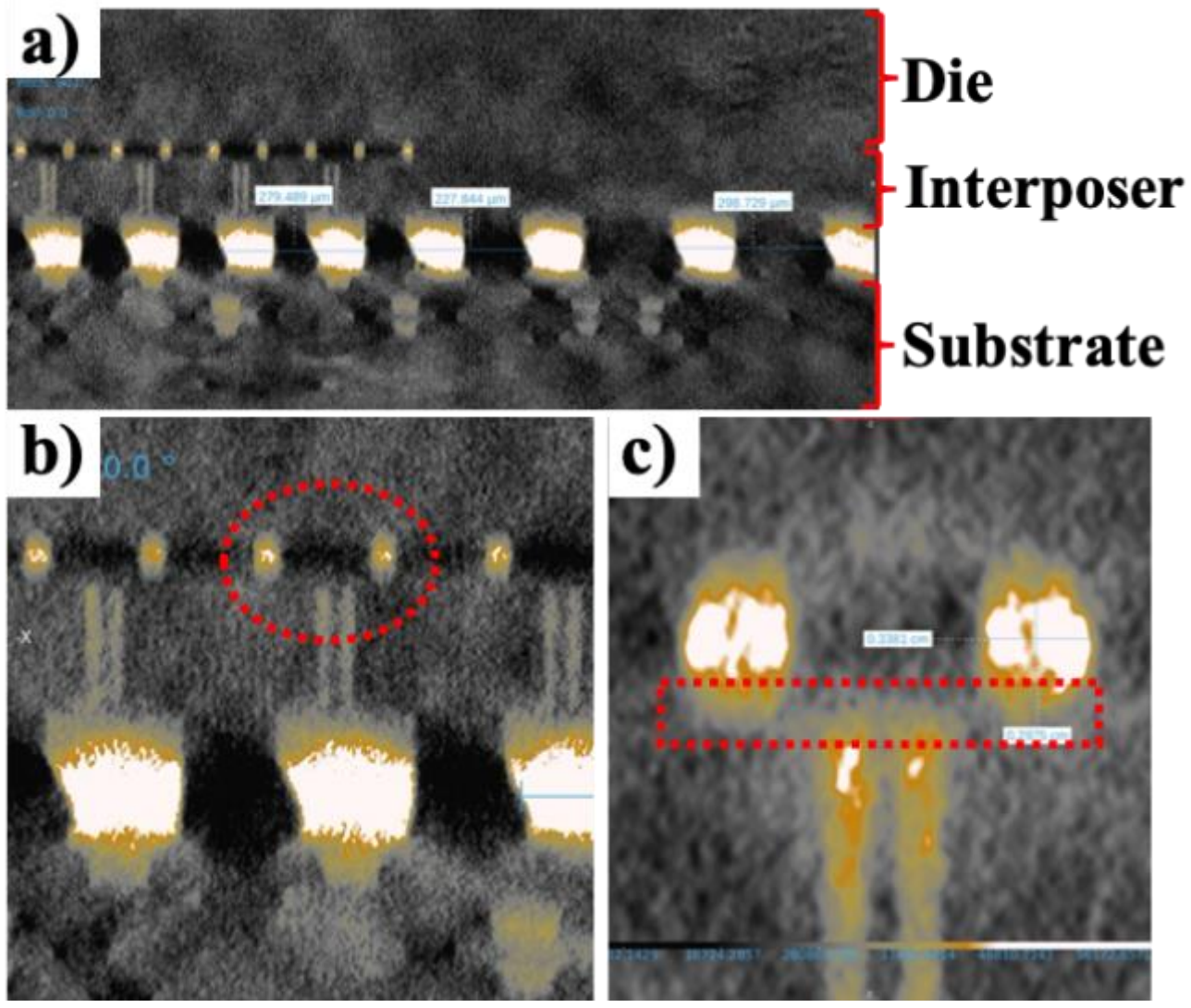

*Figure 2: a) Overall cross section of Nvidia P100, b) interposer cross section, and c) Microbump and TSV area.*

To address X-ray inspection challenges, the concept of Design for Inspection (DFI) is introduced, proposing modifications to designs to improve inspection efficiency. DFI parameters, including both design specifications and material composition, play a crucial role in addressing issues related to X-ray noise and blur [6]. To optimize the inspection process from the design stage, it is necessary to establish design guidelines based on DFI parameters and their impact on inspection efficiency.

To address the current lack of inspection compatibility-based design guidelines in advanced packages, we propose a flexible AI-integrated Design of Experiment (DoE) framework to systematically analyze and optimize design parameters for X-ray compatibility in advanced packaging technologies. This framework examines the effects and interactions of various design parameters on X-ray inspection outcomes.

## II. Related Work

### *A. X-ray Compatibility Enhancement*

Although the study of X-ray compatibility issues is a relatively nascent field, some research efforts have been directed toward developing viable solutions. One such solution is the X-ray compatibility metric, called $CM_{X\text{-ray}}$, which quantitatively evaluates the efficiency of designs regarding X-ray inspection and aids in design optimization. This metric is created through a two-step process. First, DFI parameter thresholds are established. Second, a test package design is evaluated against these thresholds to generate a numerical value that indicates X-ray compatibility, ensuring reliable detection of fine-pitched features [6]. Also, the integration of X-ray compatibility into the design process was proposed through a framework that predicts optimal design specifications, balancing imaging quality with design constraints, performance goals, and higher component density [7]. Finally, X-ADAPT was proposed as a solution when design updates to enhance X-ray compatibility are not feasible due to stringent performance constraints. By utilizing design-based inspection, X-ADAPT improves the inspection process through a customized inspection strategy [8].

### *B. Design Guideline Development*

Design guideline development is a critical process in various domains of integrated circuit (IC) and package design, test, and validation. A commonly used set of design rules, known as DRC, ensures that physical layouts comply with semiconductor manufacturers' parameters to manage process variability and improve yield [9]. Clear and well-defined rules facilitate the automation of rule checking and design optimization; for example, a DRC rule automating and checking framework was proposed that standardizes and automates the interpretation of design rules into specific categories, using a parameterized library to accurately generate DRC code [10]. Beyond design functionality, these guidelines are vital in security applications, exemplified by frameworks like ARC-FSM-G [11], which checks security rules for Finite State Machines based on AVFSM analyses [12] and MITRE CWE list of hardware weaknesses [13]. These foundational rules not only aid in identifying security vulnerabilities but also pave the way for advancements such as large language model (LLM)-assisted vulnerability databases [14], which are crucial for AI-based rule-checking frameworks to detect and mitigate security threats in hardware designs. Additionally, design rules are also applicable in specialized fields like microwave flip chip applications [15] and frequency-coded chipless radio frequency identification (RFID) [16], where precise guidelines ensure device performance and reliability, underscoring the universal importance of meticulous design guideline development.

## III. Design Guideline Development Framework

An AI-integrated DoE-based framework is proposed to analyze the effect of each design parameter on X-ray compatibility and to study the interactions between different design parameters in X-ray compatibility experiments. This framework offers a solution that is flexible with package technology and systematically addresses the lack of X-ray compatibility-based design guidelines in advanced packaging. This framework supports the optimization of package designs to balance manufacturability, performance, and inspection reliability, addressing the inspection compatibility gap in current design methodologies.
DoE is a systematic method for studying the relationships between multiple input factors and output responses. Widely used in engineering, manufacturing, and research, DoE methods provide structured and efficient ways to explore the design space, identify significant factors, and optimize processes with minimal experimentation. AI is integrated into the DoE to enhance experimental design, data analysis, and optimization. AI techniques aid in predicting outcomes, identifying patterns, and making data-driven decisions.

### *A. Framework Workflow*

The framework's steps, as shown in Figure 3, start with problem definition, focusing on X-ray compatibility analysis. Quantitative measures of X-ray compatibility should be defined, for which metrics such as signal-to-noise ratio, contrast-to-noise ratio, and noise power spectrum are used. These metrics are combined to quantitatively assess X-ray compatibility in the design. Next, factor selection is conducted, defining variables and their levels (e.g., interposer thickness with levels: low (0-50 μm) and high (50+ μm)). Following this, we select the DoE method (e.g., Plackett-Burman Design) to identify the most significant factors with minimal runs. The experiment design includes combinations of factors and levels. Subsequently, experiments are conducted by defining these combinations and performing X-ray simulation and post-processing. Data analysis and AI integration involve using machine learning techniques to build predictive models from the collected data, employing algorithms such as regression models and neural networks. Finally, implementation involves applying validated findings and drawing conclusions about design constraints.
The conclusions drawn from the analysis of experiments based on the proposed framework help establish thresholds for each analyzed design parameter, identify the most significant design parameters affecting X-ray compatibility, and elucidate the interactions and trade-offs between different design parameters and X-ray compatibility.

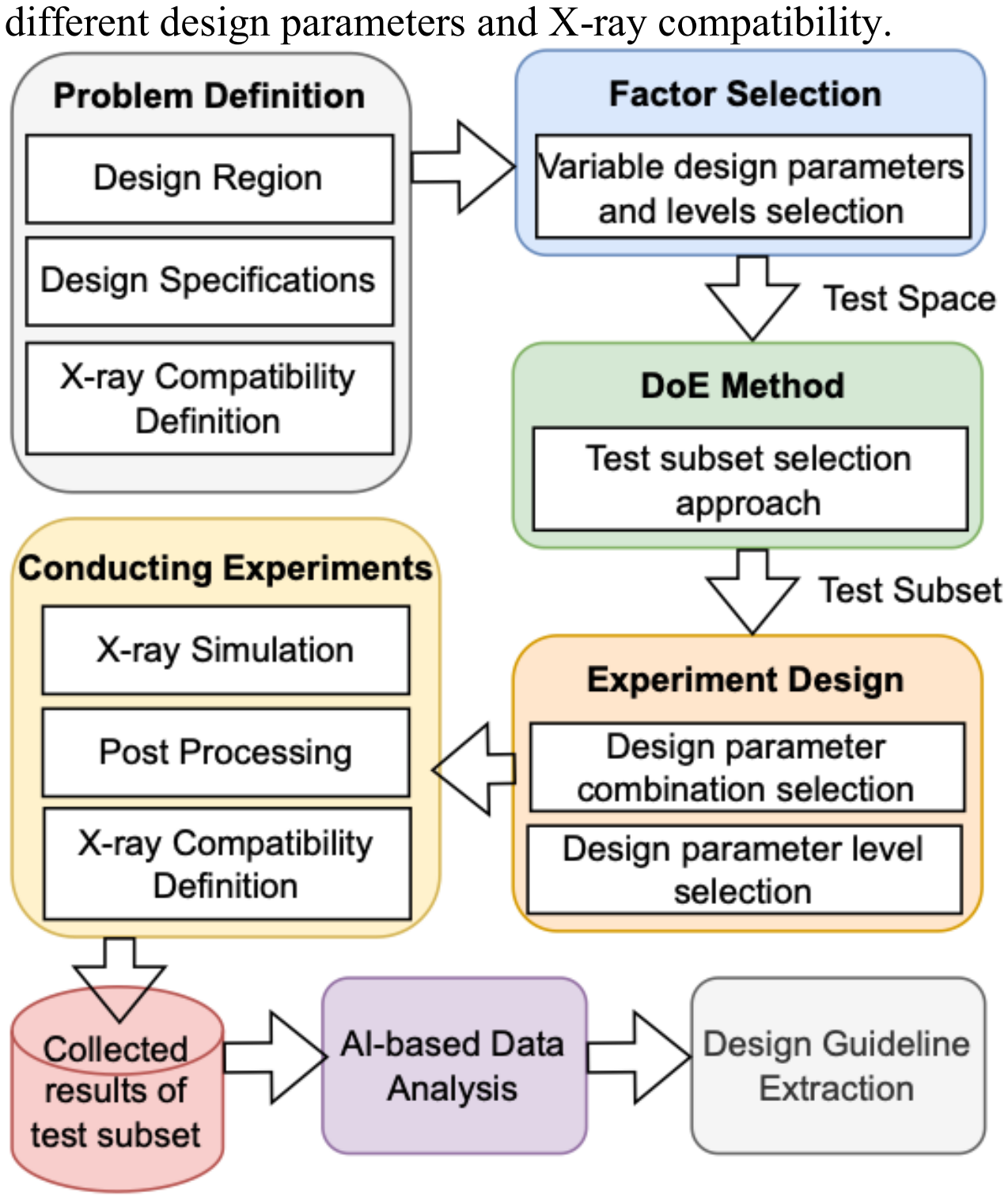

*Figure 3: Overview of proposed AI-integrated DoE-based design guideline development framework*

### *B. Design Guideline Extraction*

Main elements of design guidelines for X-ray compatibility include recommendations on component placement and orientation to avoid shadows and overlaps, and material selection to ensure optimal X-ray transparency or contrast. These guidelines also address layer thickness and profile considerations to strike a balance between structural integrity and X-ray compatibility. They provide directives for interconnect layouts to prevent the creation of X-ray opaque regions.
An efficient and flexible approach for extracting design guidelines to enhance inspection efficiency involves developing design guidelines that relate key design parameters, also known as DFI parameters, to the resolution limits of X-ray imaging systems. Table 1 presents an example of such guidelines for a typical CoWoS technology design, where thresholds for parameters such as bump pitch, bump-to-RDL spacing, RDL-to-RDL spacing, RDL interconnect pitch, RDL-to-TSV spacing, and TSV pitch are defined as multiples of the imaging system resolution *R*. These parameters are visualized in Figure 4Figure 4, where RDL-related DFI parameters are only shown in package substrate due to resolution limitations of the X-ray image.
The threshold values can be determined through rigorous empirical analysis with proposed AI-integrated DoE-based framework, ensuring that each feature remains distinguishable, and defects are detectable in X-ray scans.

The smallest component that can be seen with an X-ray imaging system, given a resolution R, is typically around the same size. However, the goal is to inspect components and detect possible defects, which are usually only a fraction of the component's size. Therefore, to achieve clear delineation and accurate interpretation, features should be 2 to 3 times larger than the resolution.

*Table 1: Example of DFI parameter design guidelines based on X-ray imaging system resolution.*

| Notation | DFI Parameter | Threshold Value (R=0.5 µm) |
|---|---|---|
| $P_b$ | Bump pitch | 50R |
| $S_{br}$ | Bump-RDL spacing | 5R |
| $S_{rr}$ | RDL-RDL spacing | 3R |
| $P_i$ | RDL interconnect pitch | 3R |
| $S_{rt}$ | RDL-TSV spacing | 4R |
| $P_t$ | TSV pitch | 75R |

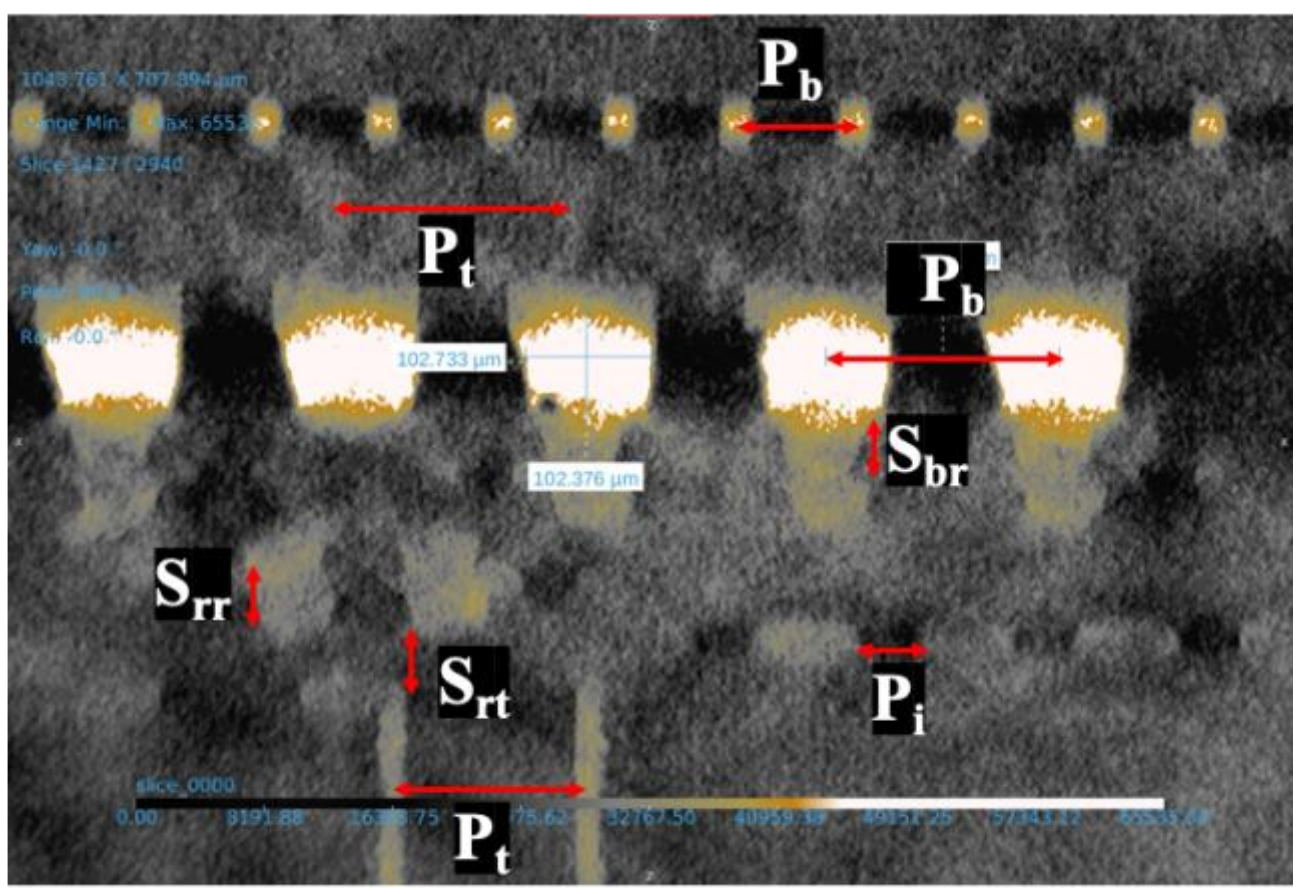

*Figure 4: Illustration of DFI parameters in 0.7µm resolution scan of Nvidia Tesla P100 sample cross section.*

The key design parameters and their respective threshold values in the example, scaled by the resolution R (where R = 0.5 µm in this context), are as follows:

- Bump pitch ($P_b$): The suggested spacing between bumps is around 50R to avoid overlap and ensure clear separation under X-ray imaging.
- Bump to RDL spacing ($S_{br}$): To maintain clarity between bumps and the underlying RDL, spacing of 5R is recommended.
- RDL to RDL spacing ($S_{rr}$): It is advised that adjacent RDL traces have a spacing of at least 3R to prevent signal interference and improve imaging resolution.
- RDL interconnect pitch ($P_i$): The pitch between interconnects within the RDL should be no less than 3R, ensuring distinguishable lines in the X-ray image.
- RDL to Through-Silicon Via (TSV) spacing ($S_{rt}$): To avoid imaging artifacts around vias, a spacing threshold of 4R is necessary.
- TSV pitch ($P_t$): TSVs must maintain a pitch of at least 75R to ensure clear detection and separation.

By anchoring design rules to the imaging system's resolution, these guidelines can offer a scalable and adaptable approach that aligns physical layout parameters with inspection capabilities. Implementing such inspection-aware design constraints early in the development cycle can significantly enhance inspection reliability, reduce costly redesign iterations, and enable high-throughput inspection in complex advanced packaging environments.

## IV. Experimental Results

A CoWoS case study is conducted to demonstrate the effectiveness of the proposed framework, elaborate on its workflow, and examine the design rules and guidelines for this widely used packaging technology that can serve as a reference.

### *A. Problem Definition*

The definition of the problem begins by analyzing the suggested TSMC design guidelines, components, design parameter ranges, material composition, total device size, and common applications for CoWoS packaging technology. Next, real X-ray images are examined in terms of X-ray compatibility to identify challenging design areas. For this case study, X-ray images of the Nvidia P100 were carefully investigated, and the interposer area, especially around the microbumps and the RDLs directly beneath them, was identified as the most challenging region for X-ray inspection. The design region model used in the X-ADAPAT study [8] is also employed as an example here. Consequently, the considered case study, shown in Figure 5, is a small area between the die and the C4 bump, including key CoWoS design components around the interposer area. The design specifications and X-ray compatibility definition from the X-ADAPAT study [8] are also reused in this analysis. A defect is inserted in one of the microbumps in the case study model and the design compatibility is defined by the distinguishability of the defective microbump and normal ones.

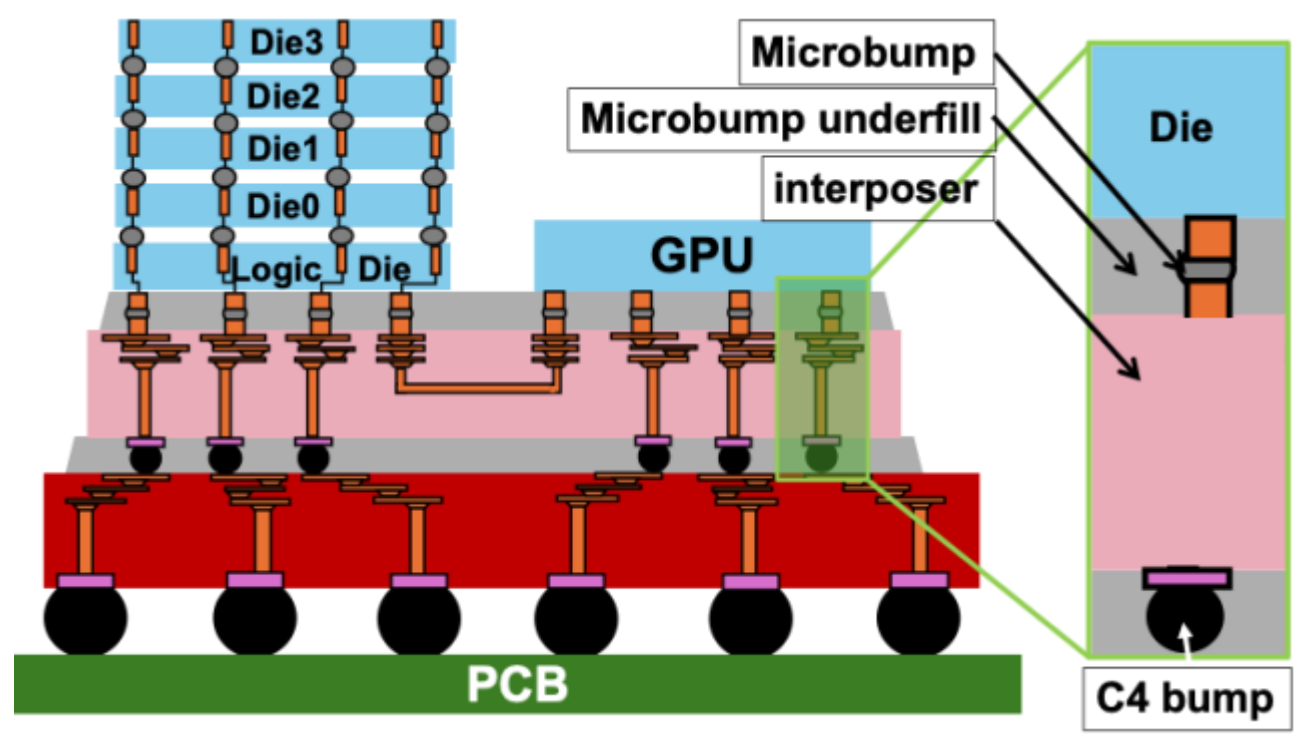

*Figure 5: Location of case study design region in a 2.5D CoWoS package structure.*

### B. Factor Selection

All the design parameters and specifications in the defined design region should be listed. The parameters for this case study include: microbump and C4 bump pitch, height, width, and underfill density and atomic number; interposer thickness, material density and atomic number; and die thickness and material properties. The factor selection stage involves selecting the variable parameters and their levels (value ranges) for performing experiments. For simplicity and to maintain the focus on the design guideline development task, we consider the variable parameters and their ranges to be similar to those in the X-ADAPT case study [8]. These parameters are microbump pitch, height, underfill density, and interposer thickness, with the assumption that the microbumps are perfect spheres, meaning their height and width are the same. All the design parameters in the case study with their corresponding values are listed in Table 2. The selection of a limited number of variable parameters from all the existing design parameters, constrains the problem, allowing it to converge to an optimum faster and with fewer experiments. The choice of variable parameters can be informed by real X-ray scan investigations and prior X-ray compatibility issues studies.

*Table 2: Case study design parameters and their values.*

| Design Parameters | Variable/ Constant | Values |
|---|---|---|
| Microbump pitch | Variable | 20 to 40 @ step of 5 µm |
| Microbump height | Variable | 12 to 27 @ step of 5 µm |
| Microbump underfill density | Variable | 1.8 to 2 @ step of 0.1 $g/cm^3$ |
| Interposer thickness | Variable | 50 to 100 @ step of 10 µm |
| Die thickness | Constant | 700 µm |
| C4 bump pitch | Constant | 250 µm |
| C4 bumps material | Constant | SnAg |
| C4 height | Constant | 100 µm |
| C4 underfill density | Constant | 2.5 $g/cm^3$ |

### C. Experiment Design

The factor selection determines the test space, while choosing a subset of this space to obtain generalized and conclusive results regarding design guidelines depends on the sampling approach (i.e., the DoE method), which is a part of experiment design. In this study, a hierarchical partitioning scheme, proposed in the context of an optimization problem that utilizes a two-stage Bayesian optimization (TSBO) algorithm [17], is used to select samples from the test space. This method is particularly suited to addressing challenges encountered in Electronic Design Automation (EDA) problems. Figure 6 illustrates an example of a partitioning tree constructed using TSBO with 4 variable parameters (d=4), showing sampled points (where the branch is expanded) and candidate points (where the branch is not yet expanded). In the proposed TSBO-based hierarchical partitioning trees, the selected branch is not fully expanded (meaning not all child nodes are sampled). The scheme expands the selected branch to generate $2^d$-1 candidate points (child nodes), but sampling occurs only at the most promising child node, as determined by the TSBO algorithm.

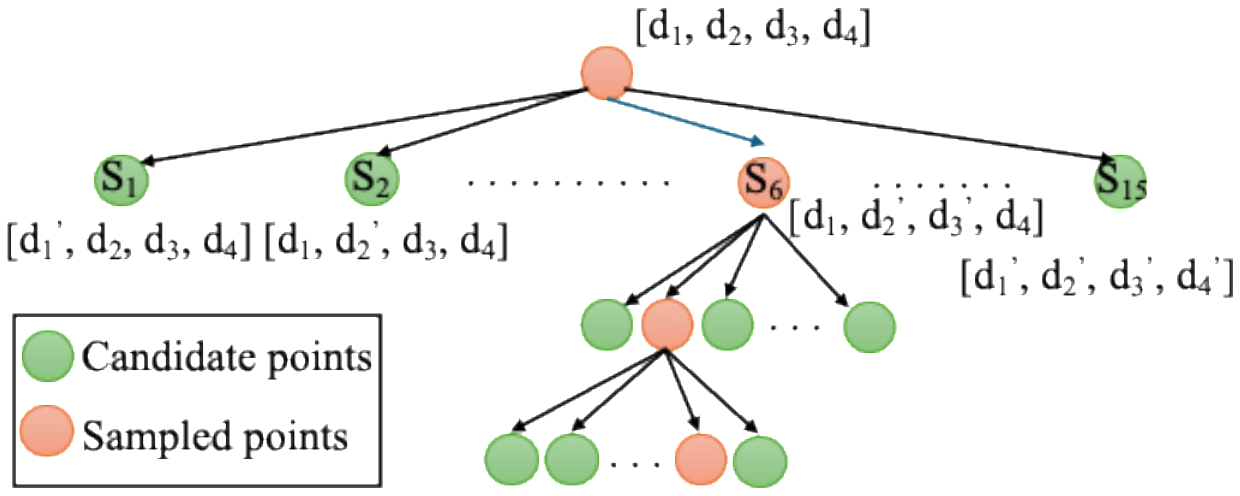

*Figure 6: Example of TSBO-based partitioning tree for sampling.*

### D. Conducting Experiments

The framework's need to analyze different sets of design parameters is accommodated by the flexibility provided by X-ray simulation, through NoviSim X-ray simulation software [18]. The simulated images are subsequently processed so the X-ray compatibility quantification metric can be obtained.

The imaging setup parameters need to be specified for X-ray simulations. Since we aim to extract design guidelines that optimize the observability of the design during in-line inspection, the X-ray system setup in the experiments should be as close as possible to real in-line inspection conditions. Some of the most important imaging setup parameters are listed in Table 3**Error! Reference source not found.**. The resolution of the X-ray imaging system can be subsequently calculated to be around 7 µm. One limitation of X-ray inspection for advanced packages is the source-object distance, which limits resolution due to the large sample sizes of packages with heterogenous integration technology. Although the design region in this case study is small, the average size of a whole sample was considered for 360-degree rotation. Also, to emphasize X-ray inspection limitations, the source-sample distance was set to a relatively large value.

*Table 3: X-ray imaging setup in X-ray simulation.*

| Imaging Setup Parameters | Values |
|---|---|
| Source-sample distance | 15 cm |
| Sample-detector distance | 35 cm |
| Region of interest size | 1024 × 512 pixels |
| Source spot dimensions | 10 µm × 10 µm |
| Detector pixel pitch | 3 µm × 1 µm |
| Exposure time | 1 s |
| Rotation | 0-360° |
| Number of projections | 10 |

### E. AI-based Data Analysis

After conducting the experiments and collecting the results, we analyzed the data to identify patterns, including the most significant input parameters and the relationship between changes in one parameter and changes in overall X-ray compatibility. To identify the most influential input parameters, we applied machine learning to quantify variable importance and visualized the resulting importance scores.
Approximately 40 X-ray simulation and compatibility evaluation experiments are conducted for the case study design region, which accounts for about 15% of the test space. The collected data is used to train a random forest (RF) AI-based classifier model to perform a binary classification of the X-ray compatibility evaluation metric, which is labeled 0 or 1 based on a predefined threshold value. Multiple iterations with different train-test splits and sampling are performed to prevent model overfitting, and the average feature importance for each iteration is recorded. Figure 7 shows the average feature importance of the four considered variable parameters in the case study. RF is selected for feature importance analysis for two reasons. First, in a prior evaluation on an analogous dataset, Logistic Regression, Naïve Bayes, Gaussian Process Classification, and Random Forest were compared; Random Forest achieved the best overall performance (accuracy and log-loss) with minimal tuning [8]. Second, tree ensembles provide native, stable global importance estimates by averaging over many decorrelated trees and can capture nonlinear effects and interactions expected in this problem [19]. Although model-agnostic explainers (e.g., permutation or SHAP [20]) can be applied to other models, with a small sample size they tend to exhibit higher variance and add tuning complexity. RF therefore offered the best trade-off between small-sample robustness and interpretable variable-importance for this study.

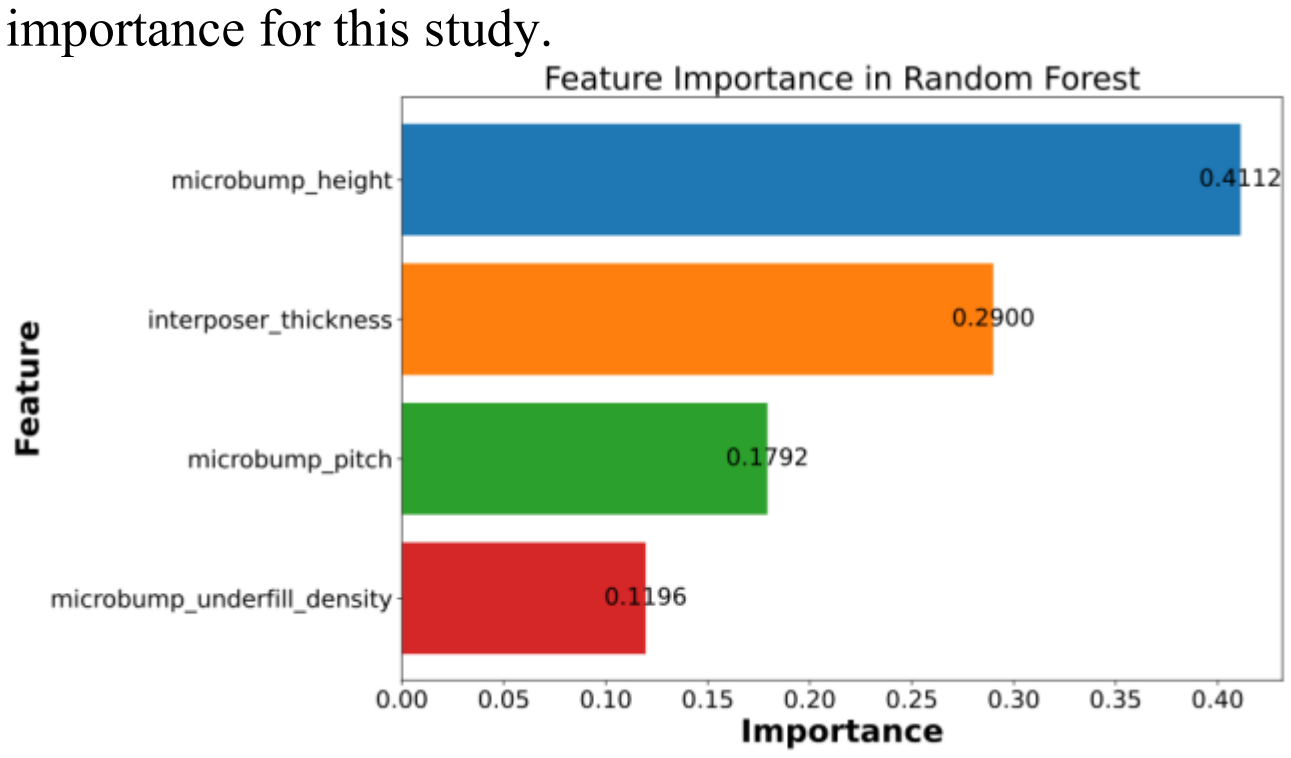

*Figure 7: Feature importance visualization of the case study design, as a result of analysis with a random forest model.*

The feature importance results indicate that microbump height, which represents microbump size, is the most significant feature in determining X-ray compatibility. This finding is predictable since defined X-ray compatibility metric focuses on microbumps and their defect detectability, making microbump size a crucial factor. The second notable feature is the interposer thickness. Being a relatively dense material as it can be simplified to a Silicon body, interposer's thickness greatly affects X-ray blur and imaging artifacts, making it an important parameter. Microbump pitch and microbump underfill density are ranked third and fourth, respectively, in feature importance. Considering the significance of microbump size in X-ray compatibility, this ranking is justified. Microbump pitch and size are closely related, as pitch is dependent on size. Also, the relatively lower density of microbump underfill in comparison to neighboring materials explains its low feature importance.

## V. Discussion

The ongoing evolution of interposer and substrate technologies, characterized by the shift from passive to active interposers, the adoption of finer RDLs, and the integration of hybrid bonding techniques, introduces new complexities that significantly impact the development of design constraints for X-ray compatibility in advanced IC packages. As packaging architectures grow increasingly dense and heterogeneous, these advances bring both opportunities and challenges for reliable inspection and defect detection via X-ray inspection.
The challenges stemming from the shrinking of feature sizes and component density increase have been investigated. Compounding these issues is the growing diversity of materials used in modern substrates and interposers. The introduction of organic composites, glass-core substrates, and silicon interposers results in variable X-ray attenuation properties that can create inconsistent image contrast and elevated noise levels. Such variability directly affects defect visibility and may prolong inspection times. This highlights the need for design guidelines that incorporate material-specific considerations to maintain inspection robustness. Another important aspect to consider is the fact that as packaging technologies and architectures continue to advance, the relevant design variables themselves will inevitably change. This dynamic evolution necessitates that design guideline development frameworks remain flexible and continuously updated to reflect emerging package structures, materials, and inspection technologies. Static or rigid design rules risk becoming obsolete as new challenges and parameters arise. Emerging packaging innovations, such as the integration of embedded diagnostic features within interposer structures, offer promising routes for early failure detection and enhanced reliability. However, the inclusion of these diagnostics requires careful coordination with X-ray inspection protocols to prevent interference with image quality and ensure inspection accuracy.

## VI. Conclusion

This study highlights the significant challenge of inspection incompatibility in the manufacturing of advanced packages and demonstrates the importance of DFI in addressing these issues. By establishing design constraints and guidelines, the efficiency of in-line inspection can be optimized. We proposed an AI-integrated DoE framework to develop such guidelines and presented a case study on a CoWoS technology design region. Through systematic analysis of various design specifications, we evaluated the X-ray compatibility of the design region and identified key factors influencing inspection accuracy. This work serves as a foundational step towards developing comprehensive design guidelines for advanced packaging technologies with improved inspection capabilities.